# JOINT MULTIFRACTAL ANALYSIS OF AIR TEMPERATURE, RELATIVE HUMIDITY AND REFERENCE EVAPOTRANSPIRATION IN THE MIDDLE ZONE OF THE GUADALQUIVIR RIVER VALLEY.

Ariza-Villaverde, A.B.[1]* (orcid.org/0000-0002-8549-2774), Pavón-Domínguez, P.[2] (orcid.org/0000-0002-2913-6492), Carmona-Cabezas, R.[1] (orcid.org/0000-0001-8324-4489), Gutiérrez de Ravé, E.[1] (orcid.org/0000-0002-2091-6708), Jiménez-Hornero, F.J.[1] (orcid.org/0000-0003-4498-8797).

[1]Complex Geometry, Patterns and Scaling in Natural and Human Phenomena Research Group, University of Córdoba, Gregor Mendel Building (3rd floor), Campus of Rabanales, 14071 Córdoba, Spain.

[2] Complex Geometry, Patterns and Scaling in Natural and Human Phenomena Research Group, University of Cádiz, University Avenue, Puerto Real University Campus, 11519 Spain

* Corresponding author. Tel.: +34 957 218309; fax: +34 957 218455.

E-mail addresses: g82arvia@uco.es (A.B. Ariza-Villaverde), pablo.pavon@gm.uca.es (Pavón-Domínguez, P), f12carcr@uco.es (Carmona-Cabezas, R.), eduardo@uco.es (E. Gutiérrez De Ravé), fjhornero@uco.es (F.J. Jiménez-Hornero).

**Highlights**

- The study climatic variables have multifractal nature.
- Multifractal analysis is able to study three climatic variables acting in concert.
- The most frequents cases have regular singularities.

## ABSTRACT

Previous works have analysed the relationship existing between reference evapotranspiration ($ET_0$) and other climatic variables under a one-at-a-time perturbation condition. However, due to the physical relationships between these climatic variables is advisable to study their joint influence on $ET_0$. The box-counting joint multifractal algorithm describes the relations between variables using relevant information extracted from the data singularities. This work investigated the use of this algorithm to describe the simultaneous behaviour of $ET_0$, calculated by means of Penman–Monteith (PM) equation, and the two main climatic variables, relative humidity ($RH$) and air temperature ($T$), influencing on it in the middle zone of the Guadalquivir river valley, Andalusia, southern Spain. The studied cases were grouped according to the fractal dimension values, obtained from the global multifractal analysis, which were related to their probability of occurrence. The most likely cases were linked to smooth behaviour and weak dependence between variables, both circumstances were detected in the local multifractal analysis. For these cases, the rest of Penman Monteith (PM) equation variables, neither the $T$ nor the $RH$, seemed to influence on $ET_0$ determination, especially when low $T$ values were involved. By contrast, the least frequent cases were those with variables showing high fluctuations and strong relationship between them. In these situations, when $T$ is low, the $ET_0$ is affected by the rest of PM equation variables. This fact confirmed $T$ as main driver of $ET_0$ because the higher $T$ values the lesser influence of other climate variables on $ET_0$. This condition could not be extended to $RH$ because the variability in $ET_0$ singularities was not significantly influenced by low or high values of this variable. These results show that the joint multifractal analysis can be regarded as a suitable tool for describing the complex relationship between $ET_0$, $T$ and $RH$, providing additional information to that derived from descriptive statistics.

Although, joint multifractal analysis shows some limitations when it is applied to large number of variables, the results reported are promising and suggest the convenience of exploring the relationships between $ET_0$ and other climatic variables not considered here with this framework such as wind speed and net radiation.

1. **INTRODUCTION**

Evapotranspiration (*ET*) is one of the most important components of the hydrological cycle, and its estimation is essential for scheduling irrigation systems, preparing input data for hydrological water balance models, computing actual *ET* for watersheds, regional water resource planning and analysing climate change effects (Tabari and Talaee, 2014; Tanasijevic et al., 2014; Gong et al., 2006; Villagra et al., 1995). *ET* is a combination of two separate processes whereby water is lost from the soil by evaporation and crop transpiration and evaporation. Commonly, *ET* is modelled by separating the effects of meteorological conditions from the nature of crop and soil available water content (Doorenbos and Pruit, 1977). For this reason, reference evapotranspiration ($ET_0$) was introduced to quantify the evaporative demand of the actual water-state of soil of the atmosphere. *ET* is affected by several factors, including weather parameters, crop characteristics and management and environmental aspects. However, $ET_0$ measures the evaporative demand of the atmosphere independently of crop type, crop development and management practices. As water is abundantly available at the $ET_0$ surface, soil factors do not affect it. Thus, the only factors affecting $ET_0$ are climatic parameters. Consequently, $ET_0$ is a climatic parameter and can be computed from weather data.

The most accurate manner in which $ET_0$ can be quantified is using weighing lysimeters or micrometeorological methods; however, these procedures are not practical because they are time-consuming and expensive (Gavilán et al., 2007). To this end, numerous approaches have been proposed to estimate $ET_0$, which are grouped in the following categories: i) water budgets (Guitjens, 1982), ii) combined energy and mass balance methods (Monteith, 1965; Penman, 1948), iii) temperature-based methods (Hargreaves and Samani, 1985; Blaney and Crid, 1950) iv) radiation-based methods (Priestley and Taylor, 1972;), v) mass transfer-based methods (Trabert, 1986; Papadakis, 1966; World Meteorological Organization, 1966), and vi) pan evaporation-based models (Almedeij, 2012; Liu et al., 2004). The use of one method over another is based on the number of atmospheric variables, such as air temperature, wind speed, relative air humidity and solar or net radiation, required as input. These approaches have been evaluated under different climatic conditions (Jabloun and Sahli, 2008; Berengena and Gavilán, 2005; Smith et al., 1996; Allen et al., 1989). The results of $ET_0$ estimation methods vary with climatic conditions (Eslamian et al., 2011), except for the Penman–Monteith (PM) equation, which demonstrates its superiority for estimating $ET_0$ over a wide range of climates (Jensen et al., 1990). Consequently, the PM equation is recommended as the standard equation by the United Nations Food and Agriculture Organization (FAO) and the American Society of Civil Engineers (ASCE) for estimating $ET_0$ (ASCE-EWRI, 2015). Nevertheless, a major drawback of employing the PM equation is its relatively high data demand. For the calculation of daily $ET_0$, apart from site location, the PM equation requires data for daily maximum and minimum air temperatures, relative humidity ($RH$), solar radiation and wind speed. The number of meteorological stations where all these parameters are observed is limited in many areas worldwide. The number of stations where reliable data for these parameters exist is even smaller, particularly in developing countries (Droogers and Allen, 2002).

Numerous authors have studied $ET_0$ to improve the understanding of its relationship with climatic variables. Most of these studies employed sensitivity analysis to assess the variation in $ET_0$ with climatic variables. Gong et al. (2006), explored the influence of $RH$, air temperature, shortwave radiation and wind speed on the PM equation's $ET_0$ variation in the Changjiang Basin, China. They observed that the most sensitive variable was $RH$ and the four studied climatic variables generally varied with season and region. Estévez et al. (2009) assessed the impact of climatic variables on the PM equation's $ET_0$ in southern Spain and showed a high degree of daily and seasonal variability, particularly for air temperature ($T$) and $RH$. Tabari and Talae (2014) observed the sensitivity of $ET_0$ to climate change in four types of climates and found that the sensitivity of $ET_0$ to wind speed and air temperature decreased from arid to humid climates. In all cases, sensitivity analysis was examined under a one-at-a-time perturbation condition, i.e. sensitivity analysis studies the effect of change of one factor on another (McCuen, 1973). It is well established in sensitivity studies that significant effects can be produced by a pair of variables acting in concert (Burgman et al., 1993). Such combined effects can be larger than the sum of the individual effects of two variables (Gong, et al., 2006). Nevertheless, the application of sensitivity analysis to evapotranspiration studies to date has been limited to one-at-a-time cases. In real world, the perturbation of more than one variable is likely to happen at the same time. The joint multifractal analysis studies the relations between variables when all of them coexist at the same time. The only condition considered in this regard is that the study variables have multifractal nature. Numerous authors have demonstrated the multifractal nature of $ET_0$ (Wang et al., 2014; Xie et al., 2008; Liu et al, 2006), describing the behaviour of the distribution of this variable by means of this formalism. The multifractal theory (Mandelbrot, 1982; Feder, 1998) implies that the complex and heterogeneous behaviour of a self-similar measure (i.e. statistically similar on any scale) can be represented as a combination of interwoven fractal sets (Kravchenko et al, 2009), each of which is characterised by its strength singularity and fractal dimension. This approach reveals certain levels of complexities that are overlooked by traditional statistical tools and fractal analyses (Zeleke and Si, 2004). On the other hand, it transforms irregular data into a more compact form and amplifies slight differences among the variables' distribution (Lee, 2002). The main advantage of this formalism is that its parameters are independent over a range of scales. Additionally, no assumption that the data follow any specific distribution is required. An extension of this procedure is joint multifractal theory, which is proposed by Meneveau et al. (1990). This approach examines the correlations of several multifractal variables that coexist at the same time and quantifies the relations between the studied measures. Joint multifractal analysis has previously been successfully employed to determine the relation between two variables in fields, including soils (Li et al., 2011; Zeleke and Si, 2004, 2005;), pollution (Jiménez-Hornero et al., 2011, 2010) and agronomy (Zeleke and Si, 2004; Kravchenko et al., 2000). This methodology can be extended to study the relationships between three variables. Meneveau et al. (1990) reported that the extension of this method to more than two multifractal distributions is straightforward. Considering the promising possibilities of this formalism, this research explores the simultaneous behaviour of climatic variables, $RH$, $T$ and $ET_0$ in the middle zone of Guadalquivir valley, southern Spain. Regarding the semi-arid regions of southern Spain, these climatic variables are strongly related to each other, as reported by Estévez et al. (2009).

## 2. STUDY AREA

This study was conducted in the middle zone of the Guadalquivir river valley located in Córdoba province, Andalusia, southern Spain (Fig. 1). Specifically, the data were collected in an experimental station located in the vicinity of Córdoba city, (37°51′34.9″ latitude and 04°47′48.9″ longitude; altitude: 110 m). According to the Köppen climate classification, the study area's climate is defined as Csa, with a warm average temperature and dry and hot summers. The annual mean temperature is 17.8°C, and the annual average rainfall is 621 mm. The weather conditions considerably vary year on year. Moreover, a continental effect is reflected in specific thermal and rainfall regimes: low rainfall, low *RH* and wide ranges for daily and annual temperature are distinctive of the study area (Domínguez-Bascón, 1983).

## 3. MATERIALS AND METHODS

### 3.1. Materials

The relationships between three climatic variables were studied in this work: *T*, *RH* and $ET_0$. To conduct this analysis, the times-series data were collected for 17 years (2001–2017) from the daily database of the Agroclimatic Information Network of Andalusia (Spain), belonging to the Ministry of Agriculture, Livestock, Fisheries and Sustainable Development. The agroclimatic station are situated on the latitude 37º 51' 25" N, longitude 04º 48' 10" W and altitude 117 above mean sea level (Cordoba, Andalusia). Temperature-humidity probe and wind sensors are placed 1.5 and 2 m above the soil surface respectively.

The three times-series datasets corresponded to daily time resolution, with a total length of 6205 data, as shown in Fig. 2. These time series had 3% of missing data due to maintenance problems or erroneous records. Days for which observations were not available were excluded in the analysis, as Gavilán et al. (2007) and Espadafor et al. (2011) did in their studies. Gaps were no longer than 3 days in a row. The corresponding averages, coefficients of variation (*CV*), skewness and kurtosis are listed in Table 1. The mean value of $ET_0$ was 3.47 mm/day, with the maximum and minimum values of 0.1 and 5.80 mm/day, respectively, and *CV* of 0.79. These variations were notably influenced by *T* and *RH*, as can be seen in Fig. 3, where the Pearson's coefficient between *T* and $ET_0$ is greater than 0.427 at a significance level of 0.01. For the case of $ET_0$ and $RH$, they are inversely correlated to a high Pearson's coefficient (−0.694).

## 3.2. Methodology

### 3.2.1. Joint Multifractal Analysis

This formalism was employed herein to describe the relationship between $T$, $RH$ and $ET_0$. Proposed by Meneveau et al. (1990), joint multifractal formalism is based on strange attractor formalism (Halsey et al., 1986; Grassberger, 1983; Hentschel and Procaccia, 1983), which deals with the fractal dimensions of geometric sets associated with the singularities of the measures. In this study, the proposed approach was extended to study the relationships between $ET_0$, T and $RH$. Thus, to employ this formalism, the time series of $ET_0$, T and $RH$ were divided into non-overlapping intervals of an initial time resolution, $n_{ini}$ ($\delta_{ini} = 2^1$, data = 1 day), in such a way that all of them contained at least one sample of the measure. Thus, the measures $\left(ET_{0_{ini}}\right)_j$, $\left(T_{ini}\right)_j$ and $\left(RH_{ini}\right)_j$ in any initial interval $j$ were set to be equal to the sample measurement or to the average when there was more than one sample. Subsequently, the domain was successively divided into $n$ non-overlapping intervals for a time resolution ranging from $\delta_{ini}$ to $\delta_{max}$. When the analysed time series was split into $n$ non-overlapping intervals of time resolution $\delta > \delta_{ini}$, the probability mass functions $c_{iET_0}(\delta)$, $c_{iT}(\delta)$ and $c_{iRH}(\delta)$ were defined in each grid size $i$ as follows:

$$c_{i[ET_0]}(\delta) = \frac{ET_{0_i}}{\sum_{j=1}^{n_{ini}}(ET_{0_{ini}})_j}$$

$$c_{i[T]}(\delta) = \frac{T_{ini}}{\sum_{j=1}^{n_{ini}}(T_{ini})_j} \qquad (1)$$

$$c_{i[RH]}(\delta) = \frac{RH_i}{\sum_{j=1}^{n_{ini}}(RH_{ini})_j}$$

where $ET_{0_i}$, $T_i$ and $RH_i$ were calculated as the sum of the $(ET_{0_{ini}})_j$, $(T)_j$ and $(RH)_j$ values, respectively, and included in the interval $i$ for a specific time resolution ($\delta$). The distribution of the probability mass function was analysed using the method of moments (Evertsz and Mandelbrot, 1992), and the joint partition function $\chi(q_{ET_0}, q_T, q_{RH}, )$, where $q_{ET_0}$, $_T$ and $q_{RH}$ are the statistical moments for $ET_0$, $T$ and $RH$, was calculated from $c_{iET_0}(\delta)$, $c_{iT}(\delta)$ and $c_{iRH}(\delta)$ as follows:

$$\chi(q_{ET_0}, q_T, q_{RH}, \delta) = \sum_{i=1}^{n} \left[c_{i_{ET_0}}(\delta)\right]^{q_{ET_0}} \left[c_{i_T}(\delta)\right]^{q_T} \left[c_{i_{RH}}(\delta)\right]^{q_{RH}}, \qquad (2)$$

The three exponents $(q_{ET_0}, q_T, q_{RH})$ ranged from −5 to 5 at 0.25 increments. The higher positive values of the statistical moments amplified the greater values of the time series $i$, while the higher negative values of statistical moments amplified the lower values of the time series $i$. This study considered this range of $q$ values to avoid instabilities in the multifractal analysis because higher and lower moment orders might magnify the influence of outliers in the measurements as stated by Zeleke and Si (2005).

The joint partition function has the following scaling property for multifractal measures:

$$\chi(q_{ET_0}, q_T, q_{RH}, \delta) \approx \delta^{\tau(q_{ET_0}, q_T, q_{RH})}, \qquad (3)$$

where $\tau(q_{ET_0}, q_T, q_{RH})$ is known as the joint mass exponent function and depends only on the exponents $q_{ET_0}$, $q_T$ and $q_{RH}$. This exponent can be obtained from the slope of the linear segment of a log–log plot of $(q_{ET_O}, q_T, q_{RH}, \delta)$ vs. $\delta$. At $q_{ET_0} > 0$, $q_T > 0$ and $q_{RH} > 0$, the value of the joint partition function was mainly determined by the high values of $ET_0$, $T$ and $RH$, while the influence of the low $ET_0$, $T$ and $RH$ contributed mostly to the partition function for $q_{ET_0} < 0$, $q_{T_1} < 0$ and $q_{RH} < 0$. The linear fits from the logarithmic plots of the partition function versus the time resolution show the range of temporal scales, $\delta_{min} - \delta_{max}$, for which the multifractal nature is exhibited.

The Hölder exponents $\alpha_{ET_0}$, $\alpha_T$ and $\alpha_{RH}$, characterised the singularities contained in the signal, which were inversely proportional to the strength of the singularity that described an abrupt change in the time series (Hampson and Malen, 2011). $\alpha_{ET_0}$, $\alpha_T$ and $\alpha_{RH}$ are also known as local fractal dimensions and can be determined by Legendre transformations of the $\tau(q_{ET_0}, q_T, q_{RH})$ function (Kravchenko et al., 2000):

$$\begin{aligned}
\alpha_{ET_O}(q_{ET_0}, q_T, q_{RH}) &= d\tau(q_{ET_0}, q_{T_1}, q_{RH})/dq_{ET_0} \\
\alpha_T(q_{ET_0}, q_{T_1}, q_{RH}) &= d\tau(q_{ET_0}, q_T, q_{RH})/dq_T \\
\alpha_{RH}(q_{ET_0}, q_T, q_{RH}) &= d\tau(q_{ET}, q_T, q_{RH})/dq_{RH}
\end{aligned} \quad (4)$$

The main purpose of any multifractal algorithm is to determine the distribution of singularities. The Hölder exponent can be mathematically defined as a point where conventional mathematical modelling breaks down (Cheng, 2007). In classical methods, a difficulty arises in the computation of the derivative of a noisy and discrete signal. These methods involve smoothing of discrete time-series data, and the gradient is then computed by differentiating the smoothed signal (Levy-Vehel and Berroir, 1993). In contrast, the multifractal approach directly uses the initial discrete time-series data and the information is straight extracted from singularities. The main advantage of this idea is that no information is lost or introduced by the smoothing process.

For time series (one-dimensional signal), when the Hölder exponent has an approximate value of 1, the measure is regular; therefore, a large change does not occur. When the Hölder exponents are less or greater than 1 ($\alpha \neq 1$), the variable in the time series shows a high gradient or discontinuity of the signal, and for $\alpha$ equal to 1 indicates smooth behaviour or considerably small activity (Levy-Vehel et al., 1992).

Let $N(\alpha_{ET_0}, \alpha_T, \alpha_{RH}, \delta)$ be the number of intervals of a given grid size ($\delta$), where a given combination of $\alpha$ values is found. Now, let us define $f(\alpha_{ET_0}, \alpha_T, \alpha_{RH})$ from the scaling relation:

$$N(\alpha_{ET_0}, \alpha_T, \alpha_{RH}, \delta) = \delta^{-f(\alpha_{ET_0}, \alpha_T, \alpha_{RH})}. \tag{5}$$

$f(\alpha_{ET_0}, \alpha_T, \alpha_{RH})$ can be considered as a fractal dimension of a set of intervals that correspond to the singularities $\alpha_{ET_0}$, $\alpha_T$ and $\alpha_{RH}$, respectively. A plot of $f(\alpha_{ET_0}, \alpha_T, \alpha_{RH})$ vs. $\alpha_{ET_0}$, $\alpha_T$ and $\alpha_{RH}$ is referred to as the joint multifractal spectrum. It can be calculated from the following equation (Meneveau et al., 1990; Chhabra et al., 1989; Chhabra and Jensen, 1989):

$$f(\alpha_{ET_0}, \alpha_T, \alpha_{RH}) = q_{ET_0}\alpha_{ET_0} + q_T\alpha_T + q_{RH}\alpha_{RH} - \tau(q_{ET_0}, q_T, q_{RH}). \tag{6}$$

The highest value of $f(\alpha_{ET_0}, \alpha_T, \alpha_{RH})$ corresponds to the fractal capacity dimension, $D_0$, that is equal to the Euclidean dimension (i.e. 1 when dealing with time series) in the box-counting joint multifractal analysis. $D_0$ is reached when $q_{ET_0}$, $q_T$ and $q_{RH}$ are equal to zero.

$f\left(\alpha_{ET_O}, \alpha_T, \alpha_{RH}\right)$ represents the frequency of the occurrence of a certain value of $\alpha_{ET_O}$, $\alpha_T$ and $\alpha_{RH}$ (Biswas et al., 2012). Generally, the joint multifractal spectrum is used to represent the joint dimensions $f\left(\alpha_{ET_O}, \alpha_T, \alpha_{RH}\right)$ of the analysed variables. When one statistical moment is equal to zero, the joint multifractal spectrum is identical to that for two variables. Likewise, a single multifractal spectrum is obtained when two statistical exponents are equal to zero at the same time.

### 3.2.2. Reference Evapotranspiration.

$ET_0$ is defined as the theoretical $ET$ from an extensive surface of actively growing green grass of uniform height completely shading the ground and not short of water, i.e. without water restrictions (Allen et al., 1998). In this study, the $ET_0$ values were calculated using the FAO-56 PM equation (Allen et al., 1998), which is a simplification of the original PM equation (Monteith, 1965). For the grass reference surface and daily time step, this equation is expressed as follows:

$$ET_0 = \frac{0.408\Delta\left(R_n - G\right) + 100/T + 273 U_2 \left(e_s - e_a\right)}{\Delta + \gamma\left(1 + 0.34 U_2\right)}, \quad (7)$$

where $ET_0$ is reference evapotranspiration (mm day$^{-1}$); $Rn$ is the net radiation at the crop surface (MJ m$^{-2}$ day$^{-1}$); $G$ is the soil heat flux (MJ m$^{-2}$ day$^{-1}$), assumed to be zero for a daily time step (Allen et al., 1998) because step soil heat flux is small compared to net radiation when the soil is covered by vegetation; $T$ is the mean daily air temperature (°C); $U_2$ is the wind speed at a height of 2 m (m s$^{-1}$); $e_s$ is the saturation vapour pressure (kPa); $e_a$ is the actual vapour pressure (kPa); ($e_s - e_a$) is the saturation vapour pressure deficit (kPa); $\Delta$ is the slope of the saturated vapour–pressure curve (kPa °C$^{-1}$) and $\gamma$ is the psychometric constant (kPa °C$^{-1}$). The unit for the coefficient 0.408 was MJ$^{-1}$ m$^2$ mm. All daily calculations were performed following the FAO-56 (Jabloun and Sahli, 2008) equation. Therefore, grass height and bulk canopy resistance were assumed to be 0.12 m and 70 m s$^{-1}$, respectively.

### 3.2.3. Seasonality.

The analysed time series exhibited seasonal fluctuations (Fig. 2) that correspond to a nonstationary process, typical behaviour for most hydro-meteorological variables at sub-annual time scales. However, the multifractal formalism applies to stationary processes (see e.g., Mandelbrot, 1982). The periodicity in hydro-meteorological time series affects their nonlinear properties and the width of the multifractal spectrum (Krzyszczak et al., 2017; Livina et al., 2011). Thus, it is necessary to filter the time series out before studying fractal properties. With this aim the seasonal decomposition procedure provided by IBM SPSS Statistics (v.25.0, IBM, 2017) was used to identify and remove the variations associated with the seasonal effects. Therefore, the original time series were decomposed into trend, seasonal and irregular components (Fig. 4).

## 4. RESULTS AND DISCUSSION

The present work carried out the study of the simultaneous behaviour of the climatic variables $T$, $RH$ and $ET_0$ using the joint multifractal analysis. This method assumes that an individual variable is multifractal. Therefore, it was necessary to check the multifractal nature of the study climatic variables before employing this algorithm. This behaviour was confirmed by previous research, which found the fluctuations of $T$, $RH$ and $ET_0$, showing self-similar structures for several ranges of time resolutions (Baranowski et al., 2015; Wang et al., 2014; Li-Hao and Zun-Ta, 2013; Xie et al., 2008; Liu et al., 2006; Király and Jánosi, 2005). Figure 5 shows the single multifractal spectra of the climatic variables considered herein at daily time scale. The shapes of these spectra (looking like inverted parabolas) evidencing their multifractal nature.

Figure 6 shows the joint multifractal spectrum obtained for the climatic variables examined in this study. From the logarithmic plots of the partition function versus the time resolution, linear fits ($R^2 > 0.999$, for all cases) were obtained for $\delta = \delta_{ini} = 8$ days to $\delta_{max} = 512$ days. Koutsoyiannis (2003) suggested that aggregation scales should range from the basic scale to a maximum value $\delta_{max}$ chosen so that sample moments can be estimated from at last 10 data values. This analysis revealed that the climatic variables exhibited multifractal nature between time resolutions ranging from 8 to 512 days.

The joint multifractal spectrum (Fig. 6) represents the relations between the Hölder exponents of the three variables $\left(\alpha_{ET_O}, \alpha_T, \alpha_{RH}\right)$. Each singularity combination corresponded to a fractal dimension value $f\left(\alpha_{ET_O}, \alpha_T, \alpha_{RH}\right)$. When the three variables were analysed, the joint multifractal spectrum was characterised by a volume. The current study investigated the relations between $T$, $RH$ and $ET_0$ under different scenarios, each of which was determined by the range of orders ($q$) selected for the statistical moments. Table 2 lists the eight studied cases. Here, $q > 0$ and $q < 0$ mean high and low variable values, respectively.

According to Fig. 7, which displays the results obtained from the global multifractal analysis, the cases listed in Table 2 can be clustered in three groups based on the median ($M_e$) value of the fractal dimension values, $f(\alpha_T, \alpha_{RH}, \alpha_{ET0})$, as it is shown in the box-whisker plots. Thus, the first group includes Cases 1, 2 and 3 which were the most frequent situations found in the time series. A second group is composed by Cases 4, 5 and 6 which exhibited a lower probability of occurrence than those contained in the first group. Finally, the last group can be set with Cases 7 and 8 which were less likely.

The results yielded from the applying the local multifractal analysis to the components of each group are shown in Fig. 8 (Cases 1, 2 and 3), Fig. 9 (Cases 4, 5 and 6) and Fig. 10 (Cases 7 and 8). Regarding Fig. 8 it can be checked that $T$, $RH$ and $ET_0$ behaviour was regular because the corresponding Hölder exponents, $\alpha_T$, $\alpha_{RH}$, $\alpha_{ET0}$, were around 1. In addition, $\alpha_T$, $α_{RH}$, $α_{ET0}$ distributions were similar revealing low association between the studied variables. Both circumstances suggested that the rest of variables included in PM equation influenced on $ET_0$ determination, especially in Case 2 where low $T$ values were involved provoking wider $ET_0$ box-whisker plot than Cases 1 and 3. The local behaviour detected is related to frequent situations according to the fractal dimensions shown in Fig. 7 for Cases 1, 2 and 3.

Box-whisker plots displayed in Fig. 9 exhibits higher association between variables for Cases 4, 5 and 6, due to the differences in $α_T$, $α_{RH}$, $α_{ET0}$ distributions, than those included in the first group. Therefore, the rest of the PM equation variables has less influence on these cases. The fluctuations reported for $T$, $RH$ and $ET_0$ were more relevant than those detected for the previous cases 1, 2 and 3 because there were some situations in which the Hölder exponents differed to 1. Cases 5 and 6, with low $T$ values, had wider $ET_0$ box-whisker plots suggesting the influence of PM equation variables as happened in Case 2. Local multifractal analysis results obtained for the second group of cases agreed with the fact of being less recurrent likely than those included in the first group (Fig. 7).

Local multifractal analysis results for Cases 7 and 8, belonging to the least frequent group (Fig. 7), are shown in Fig. 10. The $α_T$, $α_{RH}$, $α_{ET0}$ distributions exhibited more relevant differences between them compared to the cases included in the other groups. This situation is related with stronger association between $T$, $RH$ and $ET_0$ because the Hölder exponents differed to 1 for all variables. It can be verified that low $T$ values (Case 8) were linked to wider $ET_0$ box-whisker plots indicating the influence of PM equation variables as it was commented before.

A last comment about global multifractal analysis results (Fig. 7) should be done because the stronger association between $T$, $RH$ and $ET_0$ the wider range for $f(α_T, α_{RH}, α_{ET0})$ in each group of cases, i.e. higher variability probability of occurrence. This fact was more relevant when low *T* values were involved (see Case 2 in the first group; Cases 5 and 6 in the second group; and Case 8 in the last group).

## 5. CONCLUSION

Previous works suggested physical interaction between *T* and *RH* as the most relevant process influencing on $ET_0$ in locations placed in middle zone of the Guadalquivir river Valley (Andalusia, southern Spain). Joint multifractal analysis performed here confirmed this fact according to the results yielded by extracting time series information from singularities. This information allowed to group the studied cases according to their probability of occurrence determined by the fractal dimension values from the global multifractal analysis. The most likely cases were related to smooth behaviour and weak association between *T*, *RH* and $ET_0$, both circumstances detected in the local multifractal analysis. For these cases, it was suggested that the rest of PM equation variables influenced on $ET_0$ determination, especially when low *T* values were involved. By contrast, the least frequent cases were those with high *T*, *RH* and $ET_0$ fluctuations and strong relationship between them. In these situations, the other PM equation variables effects on $ET_0$ were only relevant with low *T* values again. Thus, *T* can be regarded as main driver of $ET_0$ because the higher *T* values the lesser influence of the rest of the PM equation variables acting on $ET_0$. However, for cases with low *T* values, the variability in $ET_0$ singularities was higher and not significantly being influenced by low or high *RH* values implying the action of other PM equation variables.

To date, the relationships between $ET_0$ and other climatic variables have been analysed under a one-at-a-time perturbation condition. However, this study has explored the links between *T*, *RH* and $ET_0$ acting in concert using box-counting joint multifractal analysis. The results obtained herein are promising and expand the existing description of the complex interactions between these variables. Alternative approach to descriptive statistics, such as multifractal analysis, applied to study the $ET_0$ links to climate drivers are needed due to the increase of freshwater resources demand under global climate change.

From the theoretical point of view, joint multifractal analysis can be trivially extended to the study of more than three variables. However, some limitations, related to the computational load and graphic representation of the results, prevent its use in practice. Therefore, the joint multifractal study of $ET_0$ and some variables appearing in PM equation, such as net radiation and wind velocity, is pending as future work. Thus, new insights might be found specially for situations in which air temperature is low allowing the influence of other climate drivers on $ET_0$ different from $RH$.

**ACKNOWLEDGEMENTS**

The FLAE approach for the sequence of authors is applied in this work. Authors gratefully acknowledge the support of the Andalusian Research Plan Group TEP-957 and the XXIII research program (2018) of the University of Cordoba

**REFERENCES**

Allen, R.G. Jensen, M.E., Wright, J.L. and Burman, R.D., 1989. Operational estimates of reference evapotranspiration. Agron. J. 81, 650–662.

Allen, R.G., Pereira, L.S., Raes, D. and Smith, M., 1998. Crop evapotranspiration: guidelines for computing crop water requirements, FAO Irrigation and Drainage, first ed. Food and Agriculture Organization of the United Nations, Rome, Italy.

Almedeij, J., 2012. Modeling pan evaporation for Kuwait by multiple linear regression, Sci. World J., ID 574742.

ASCE-EWRI, 2005. The ASCE standardized reference evapotranspiration equation, first ed. (Environmental and Water Resources Institute of the American Society of Civil Engineers, Task Committee on Standardization of Reference Evapotranspiration Calculation, Washington DC, USA, p. 70.

Baranowski, P., Krzyszczak, J., Slawinski, C., Hoffmann, H., Kozyra, J., Nieróbca, A., Siwek, K., and Gluz, A., 2015. Multifractal analysis of meteorological time series to assess climate impacts. Clim. Res. 65, 39–52.

Berengena, J. and Gavilán, P., 2005. Reference evapotranspiration estimation in a highly advective semiarid environment. J. Irrig. Drainage Eng-ASCE. 131(2), 147–163.

Biswas, A., Cresswell, H.P. and Bing, C.S., 2012. Application of multifractal and joint multifractal analysis in examining soil spatial variation: A review, in Fractal Analysis and Chaos in Geosciences, ed. Quadfeul S, pp. 109–138.

Blaney, H.F. and Criddle, W.D., 1950. Determining water requirements in irrigated areas from climatologically and irrigation data, USDA Soil Conservation Service SCS, p. 44.

Burgman, M.A., Ferson, S. and Akcakaya, H.R., 1993. Risk Assessment in Conservation Biology, Chapman & Hall, London.

Cheng, Q., 2007. Mapping singularities with stream sediment geochemical data for prediction of undiscovered mineral deposits in Gejiu, Yunnan Province, China. Ore. Geol. Rev. 32, 314–334.

Chhabra A.B., Meneveau, C., Jensen, R.V. and Sreenivasan, K.R., 1989. Direct determination of the f (α) singularity spectrum and its application to fully developed turbulence. Phys. Rev. A. 40, 5284–94.

Chhabra, A.B. and Jensen, R.V., 1989. Direct determination of the f(α) singularity spectrum, Phys. Rev. Lett. 62, 1327–30.

Domínguez-Bascón, P., 2002. Clima regional y microclimas urbanos en la provincial de Córdoba, Servicio de Publicaciones de la Universidad de Córdoba, Córdoba, Spain.

Doorenbos, J. and Pruitt, W.O., 1977. Crop Water Requirements. FAO Irrigation and Drainage, 24, Food and Agriculture Organization, Rome.

Droogers, P. and Allen, R.G., 2002. Estimating reference evapotranspiration under inaccurate data conditions. Irrig. Drain Syst. 16, 33–45.

Eslamian, S., Khordadi, M.J. and Abedi-Koupai, J., 2011. Effects of variations in climatic parameters on evapotranspiration in the arid and semi-arid regions. Glob. Planet. Change. 78 (3–4), 188–194.

Espadafor, M., Lorite, I.J., Gavilán, P., and Berengena, J., 2011. An analysis of the tendency of reference evapotranspiration estimates and other climate variables during the last 45 years in Southern Spain. Agric. Water Manag. 98 (6), 1045−1061.

Estévez, J., Gavilán, P. and Berengena, J., 2009. Sensitivity analysis of a Penman–Monteith type equation to estimate reference evapotranspiration in southern Spain. Hydrol. Process. 23, 3342–3353.

Evertsz, C.J.G. and Mandelbrot, B.B., 1992. Multifractal measures. Chaos and fractals: new frontiers in science, Springer, New York.

Feder, J., 1988. Fractals. New York: Plenum Press.

Gavilán, P., Berengena, J. and Allen, R.G., 2007. Measuring versus estimating net radiation and soil heat flux: Impact on Penman–Monteith reference ET estimates in semiarid regions. Agric. Water Manage. 89, 275–286.

Gong, L., Xu, C., Chen, D. and Halldin, S., 2006. Sensitivity of the Penman–Monteith reference evapotranspiration to key climatic variables in the Changjiang (Yangtze River) basin. J. Hydrol. 329, 620–629.

Grassberger, P., 1983. Generalized dimensions of strange attractors. Phys. Lett. A. 97, 227–30.

Guitjens, J.C., 1982. Models of alfalfa yield and evapotranspiration. J. Irrig. Drain Div. 108(3), 212–222.

Halsey, T.C., Jensen, M.H., Kadanoff, L.P., Procaccia, I. and Shraiman, B.I., 1986. Fractal measures and their singularities: the characterization of strange sets. Phys. Rev. 33, 1141–51.

Hampson, K.M. and Malen, E.A.H., 2011. Multifractal nature of ocular aberration dynamics of the human eye. Biomed. Opt. Express. 1, 464–477.

Hargreaves, G.H. and Samani, Z.A., 1985. Reference crop evapotranspiration from temperature. Appl. Eng. Agric. 1(2), 96–99.

Hentschel, H.E. and Procaccia, I., 1983. The infinite number of generalized dimensions of fractals and strange attractors. Phys. D. 8, 435–44.

Jabloun, M. and Sahli, A., 2008. Evaluation of FAO-56 methodology for estimating reference evapotranspiration using limited climatic data Application to Tunisia. Agr. Water Manage. 95, 707–715.

Jensen, M.E., Burman, R.D. and Allen, R.G., 1990. Evapotranspiration and irrigation water requirements, ASCE Manuals and Reports on Engineering Practice, American Society of Civil Engineers, New York, Vol. 70.

Jiménez-Hornero, F.J., Jiménez-Hornero, J.E., Gutiérrez de Ravé, E. and Pavón-Domínguez, P., 2010. Exploring the relationship between nitrogen dioxide and ground-level ozone by applying the joint multifractal analysis. Environ. Monit. Assess. 167, 675–684.

Jiménez-Hornero, F.J., Pavón-Domínguez, P., Gutiérrez de Ravé, E. and Ariza-Villaverde, A.B., 2011. Joint multifractal description of the relationship between wind patterns and land surface air temperature. Atmos. Res. 99(3–4), 366–376.

Király, A., and Jánosi, I.M., 2005. Detrended fluctuation analysis of daily temperature records: Geographic dependence over Australia. Meteorol. Atmos. Phys. 88, 119–128.

Koutsoyiannis, D., 2003. Climate change, the Hurst phenomenon, and hydrological statistics. Hydrol. Sci. J. 48 (1), 3–24.

Kravchenko, A.N., Bullock, D.G. and Boast, C.W., 2000. Joint multifractal analysis of crop yield and terrain slope. Agron. J. 92(6), 1279–1290.

Kravchenko, A.N., Martin, M.A., Smucker, A.J.M., and Rivers, M.L., 2009. Limitations in determining multifractal spectra from pore–solid soil aggregate images. Vadose Zone J. 8, 220–226.

Krzyszczak, J., Baranowski, P., Zubik, M., and Hoffmann, H., 2017. Temporal scale influence on multifractal properties of agro-meteorological time series. Agric. For. Meteorol. 239, 223–235.

Lee, C.K., 2002. Multifractals characteristics in air pollutant concentration times series. Water Air Soil Pollut. 135, 389–409.

Levy-Vehel, J. and Berroir, J.P., 1942. Image analysis through multifractal description, presented at the Fractal's 93 conference, INRIA, London, United Kingdom.

Levy-Vehel, J., Mignot, P. and Berroir, J.P., 1992. Multifractal, texture and image analysis. Computer Vision and Pattern Recognition, in Proceedings 1992 IEEE Computer Society Conference on Computer Vision and Pattern Recognition, Champaign, IL, USA.

Li, Y., Li, M. and Horton, R., 2011. Single and joint multifractal analysis of soil particle size distributions. Pedosphere. 21(1), 75–83.

Li-Hao, G. and Zun-Ta, F., 2013. Multi-fractal behaviors of relative humidity over China. Atmos. Oceanic Sci. Lett. 6(2), 74−78.

Liu, B., Shao, D. and Shen, X., 2006. Research on temporal fractal features of reference evapotranspiration. J. Irrig. Drain. 25(5), 9–13.

Liu, B., Xu, M., Henderson, M. and Gong, W., 2004. A spatial analysis of pan evaporation trends in China. J. Geophys Res. 109 (D15102), 1955–2000.

Livina, V.N., Ashkenazy, Y., Bunde, A., and Havlin, S., 2011. Seasonality effects on nonlinear properties of hydrometeorological records, in: Kropp, J. and Shellnhuber, H.J (Eds.), Springer, Berlin.

Lombardo, F., Volpi1, E., Koutsoyiannis, D. and Papalexiou, S.M., 2014. Just two moments! A cautionary note against use of high-order moments in multifractal models in hydrology. Hydrol. Earth Syst. Sci. 18, 243–255.

Mandelbrot, B.B., 1982. The Fractal Geometry of Nature, in W.H. Freeman, New York.

McCuen H.R., 1974. A sensitivity and error analysis of procedures used for estimating evapotranspiration. Water Resour. Bull. 10(3), 486–498.

Meneveau, C., Sreenivasan, K.R., Kailasnath, P. and Fan, M.S., 1990. Joint multifractal measures: theory and applications to turbulence. Phys. Rev. A. 41, 894–913.

Monteith, J.L., 1965. Evaporation and environment. The State and Movement of Water in Living Organisms, in XIX Symposia of the 1965 Society for Experimental Biology, Cambridge University Press, Swansea, UK.

Papadakis, J., 1966. Climates of the World and their Agricultural Potentialities, Papadakis, Buenos Aires.

Penman, H.L., 1948. Natural evaporation from open water, bare soil and grass, in Proceedings Royal Society A, (London, United Kingdom, Vol. 193, pp. 120–145.

Priestley, C.H.B. and Taylor, R.J., 1972. On the assessment of surface heat flux and evaporation using large-scale parameters. Mon. Weather Rev. 100, 81–92.

Smith, M., Allen, R.G. and Pereira, L.S., 1996. Revised FAO methodology for crop water requirements. Evapotranspiration and irrigation scheduling, in Proceedings International Conference, ASAE San Antonio, pp. 116–123.

Tabari, H. and Talaee, P.H., 2014. Sensitivity of evapotranspiration to climatic change in different climates. Global Planet Change. 115, 16–23.

Tanasijevic, L., Todorovic, M., Pereira, L., Pizzigalli, C. and Lionello, P., 2014. Impacts of climate change on olive crop evapotranspiration and irrigation requirements in the Mediterranean region. Agr. Water Manage. 144, 54–68.

Trabert, W., 1896. Neue Beobachtungen uber Verdampfungsgeschwindigkeiten. Meteorologische Zeitschrift, 13, 261–263.

Villagra, M.M., Bacchi, O.O.S., Tuon, R.L. and Reichardt, K.,1995. Difficulties of estimating evapotranspiration from the water balance equation. Agr. Forest Meteorol. 72, 317–325.

Wang, W., Zou, S., Luo, Z., Zhang, W., Chen, D. and Kong, J., 2014. Prediction of the reference evapotranspiration using a chaotic approach. Scientific World J. ID 347625, pp. 13.

World Meteorological Organization, 1966. Measurement and Estimation of Evaporation and Evapotranspiration, Secretariat of the World Meteorological Organization, Geneva.

Xie, X., Cui, Y. and Zhou, Y., 2008. Long-term correlation and multifractality of reference crop evapotranspiration time series. J. Hydraul. Eng. 39(12), 1327–1333.

Zeleke, T.B. and Si, B.C., 2004. Scaling properties of topographic indices and crop yield: multifractal and joint multifractal approaches. Agron. J. 96, 1082–90.

Zeleke, T.B. and Si, B.C., 2005. Scaling relationships between saturated hydraulic conductivity and soil physical properties. Soil Sci. Soc. Am. J. 69, 1691–1702.

**Figures captions**

Fig. 1. Location of the study area.

Fig. 2. Time-series data of the studied climatic variables.

Fig.3. Scatter plots and linear fits of the studied climatic variables.

Fig. 4. Decomposition of the studied variables time series into seasonal, trend and irregular components for daily series.

Fig. 5. Single multifractal spectra of the studied variables.

Fig. 6. Joint multifractal spectrum.

Fig. 7. Box–whisker plots of the fractal dimensions considered in each study case.

Fig. 8. Box–whisker plots of the singularities considered in Cases 1, 2 and 3.

Fig. 9. Box–whisker plots of the singularities considered in Cases 4, 5 and 6.

Fig. 10. Box–whisker plots of the singularities considered in Cases 7 and 8.

**Table captions**

Table 1. Statistical parameters of the studied climatic variables.

Tables 2. Descriptions of the studied scenarios.

**Table**

Tables

Table 1. Statistical parameters of the studied climatic variables.

| Climatic Variables | N | Minimum | Maximum | Average | CV | Skewness | Kurtosis |
|---|---|---|---|---|---|---|---|
| T (ºC) | 6205 | 9,2 | 26,65 | 18,31 | 2,54 | -,106 | -,123 |
| RH (%) | 6205 | 22,6 | 95,00 | 63,28 | 10,70 | -,086 | ,033 |
| $ET_O$ (mm/day) | 6205 | 0,0 | 5,80 | 3,47 | 0,79 | -,860 | 1,880 |

Tables 2. Studied sceneries description.

|        | T     | RH    | $ET_0$ |
|--------|-------|-------|-------|
| **CASE 1** | $q > 0$ | $q > 0$ | $q < 0$ |
| **CASE 2** | $q < 0$ | $q < 0$ | $q > 0$ |
| **CASE 3** | $q > 0$ | $q > 0$ | $q > 0$ |
| **CASE 4** | $q > 0$ | $q < 0$ | $q < 0$ |
| **CASE 5** | $q < 0$ | $q < 0$ | $q < 0$ |
| **CASE 6** | $q < 0$ | $q > 0$ | $q > 0$ |
| **CASE 7** | $q > 0$ | $q < 0$ | $q > 0$ |
| **CASE 8** | $q < 0$ | $q > 0$ | $q < 0$ |

$q > 0$ and $q < 0$ mean high and low variable values, respectively.

**fig.1.**

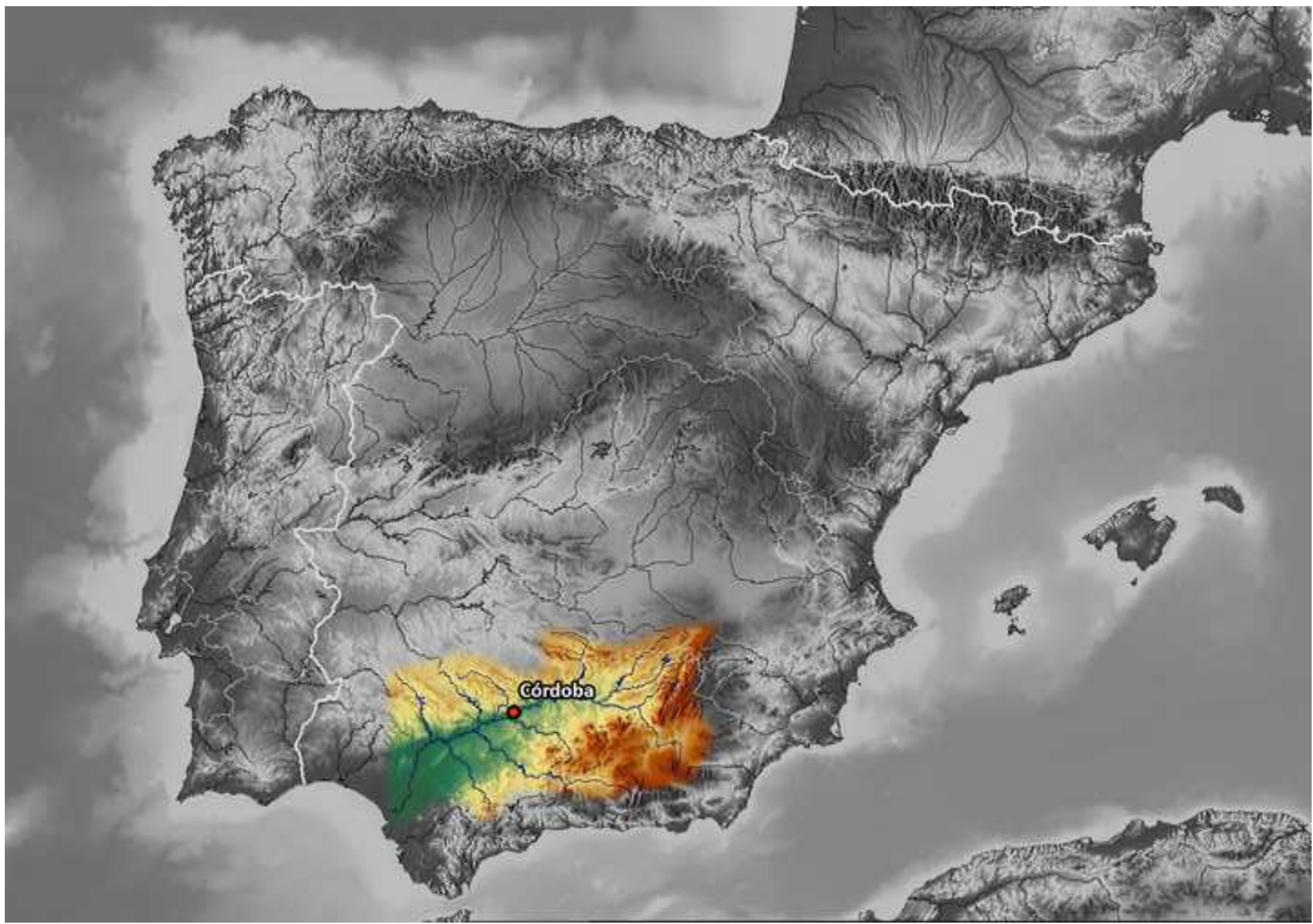

**fig.2.**

## ORIGINAL TIME SERIES

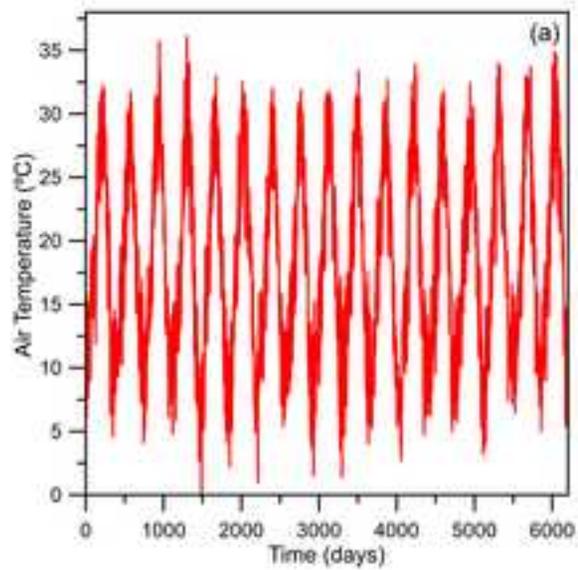 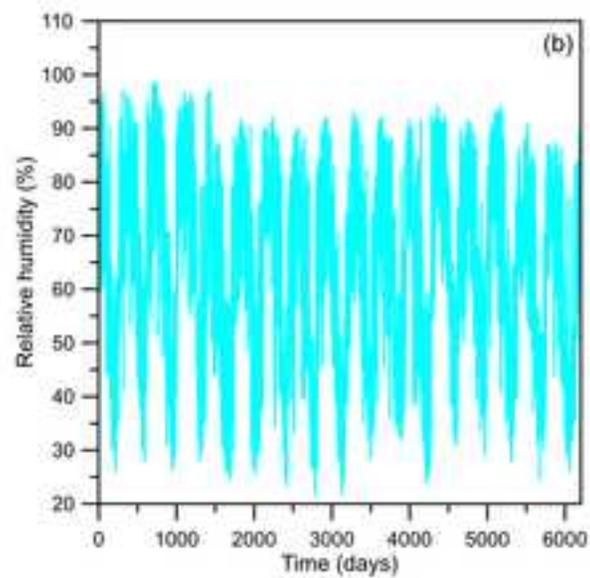 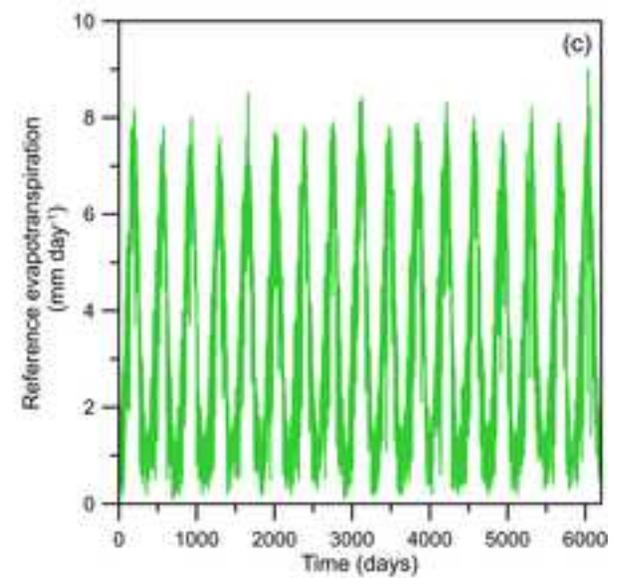

## DESEASONALIZED TIME SERIES

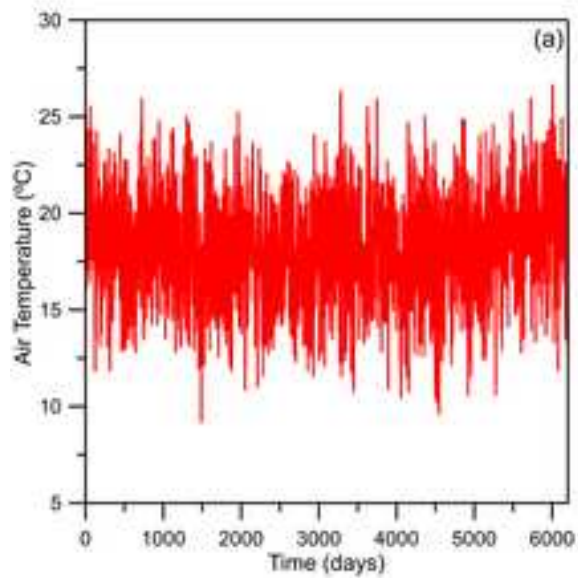 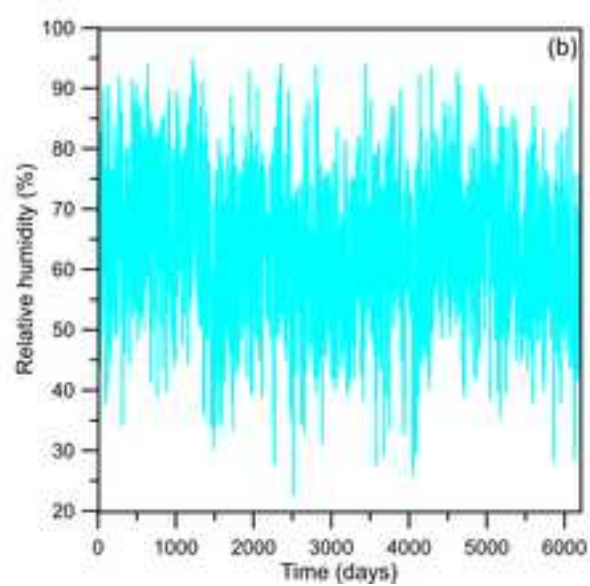 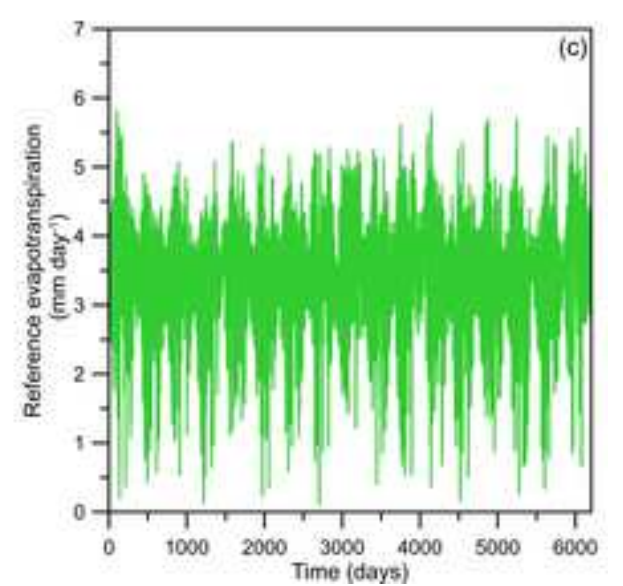

**fig.3.**

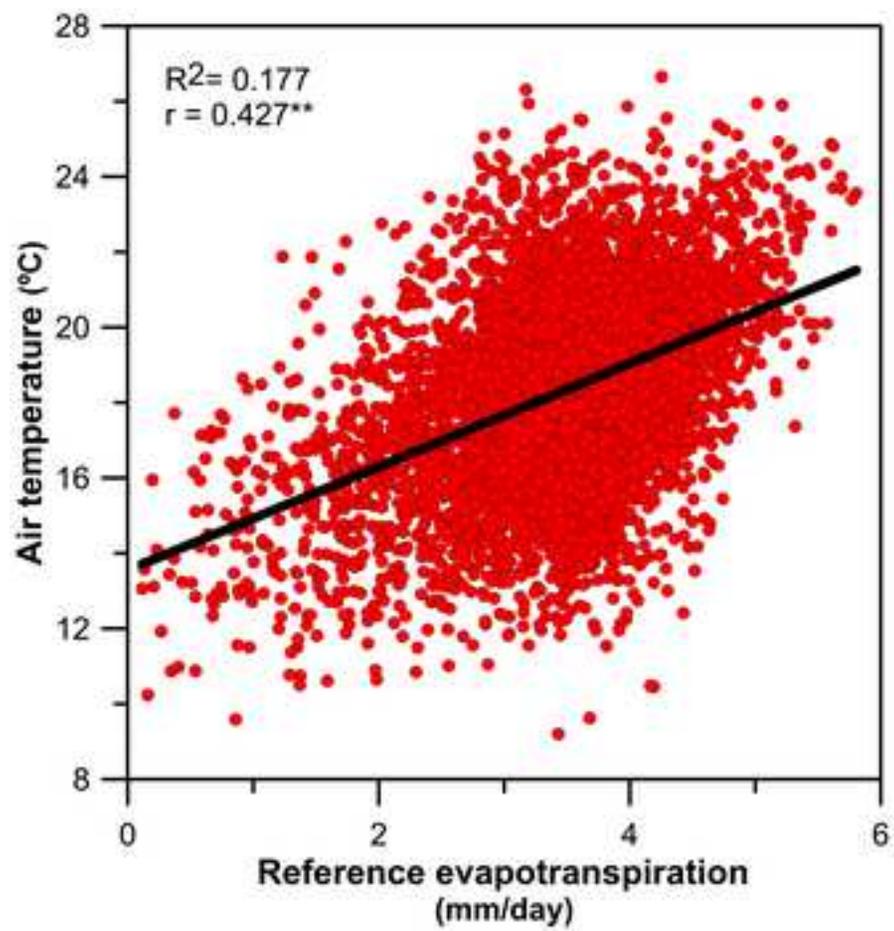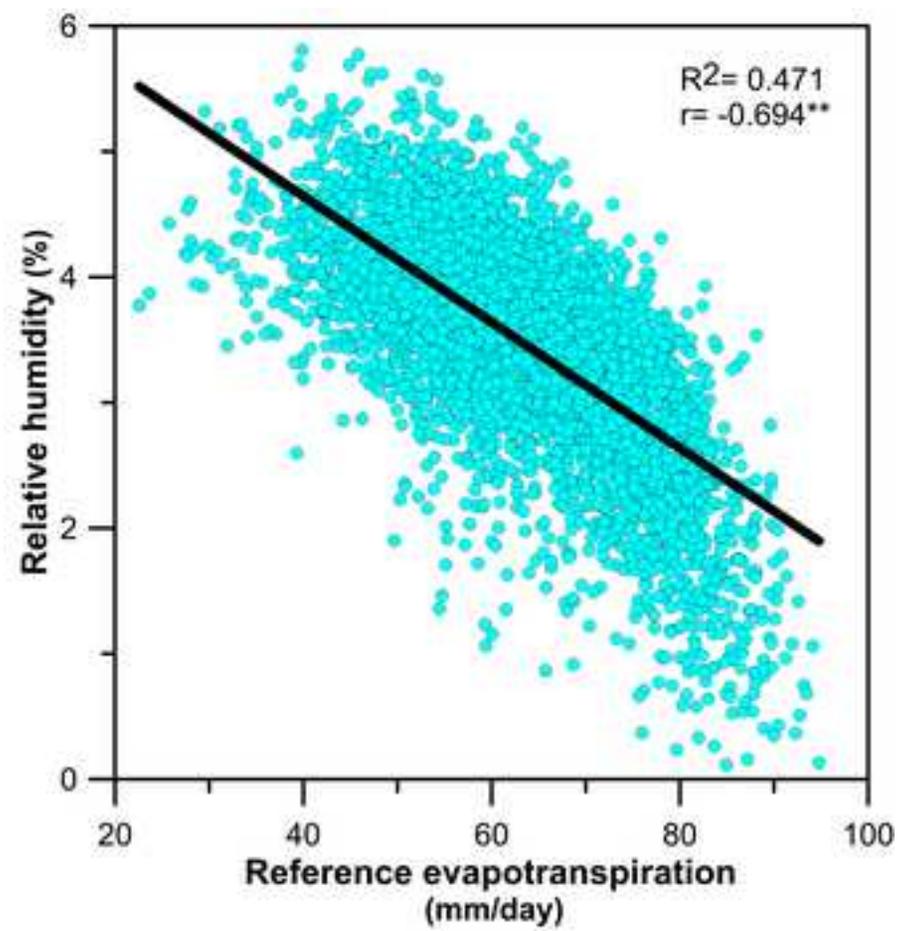

**fig. 4.**

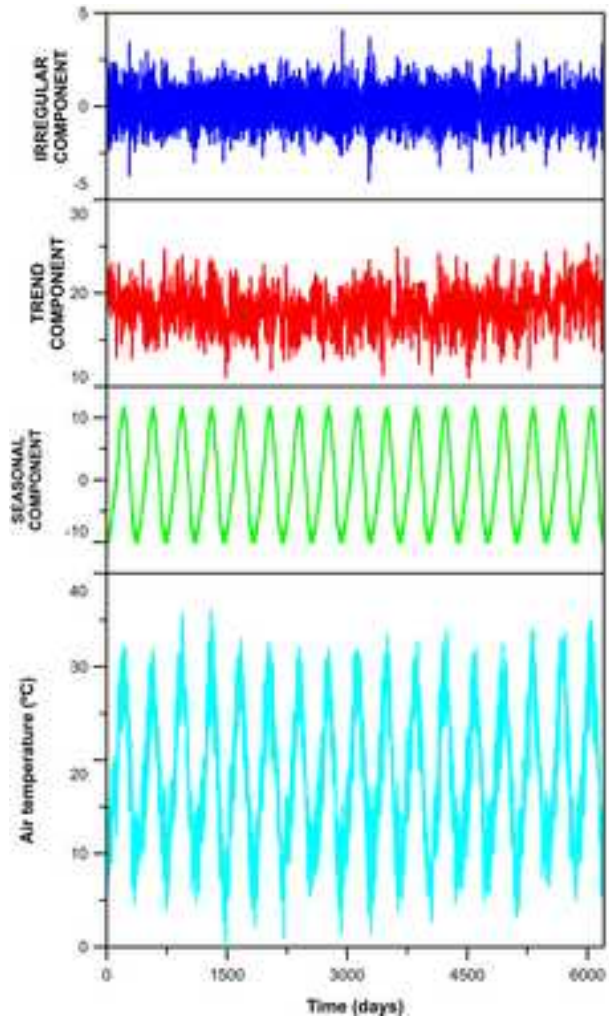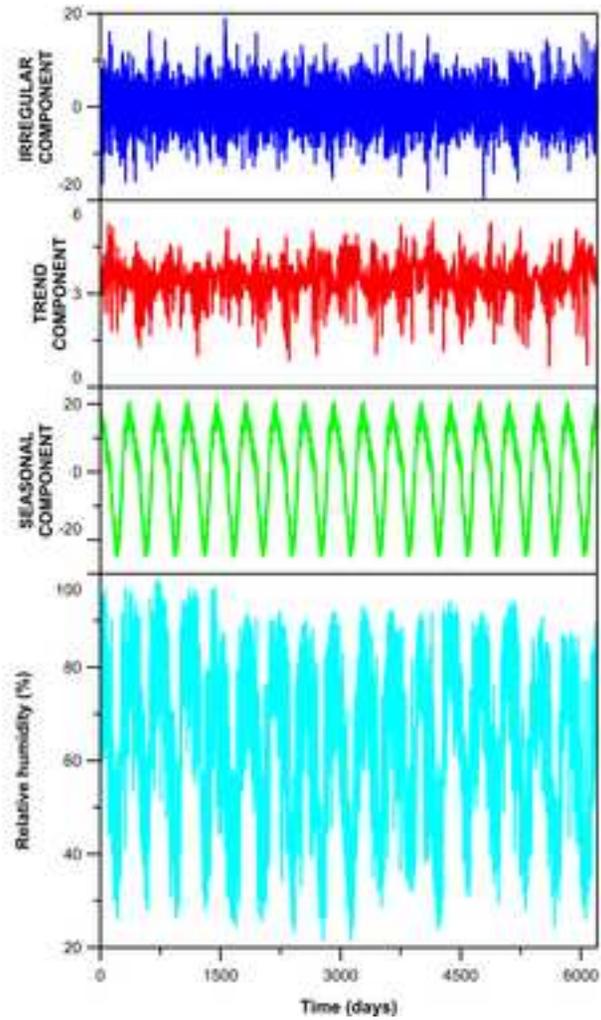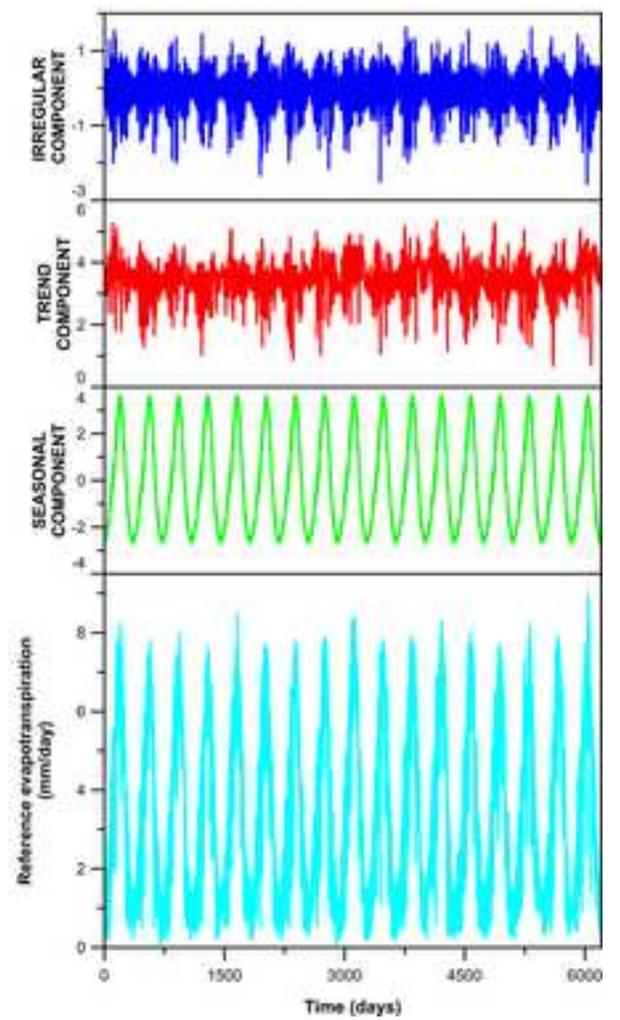

fig, 5.

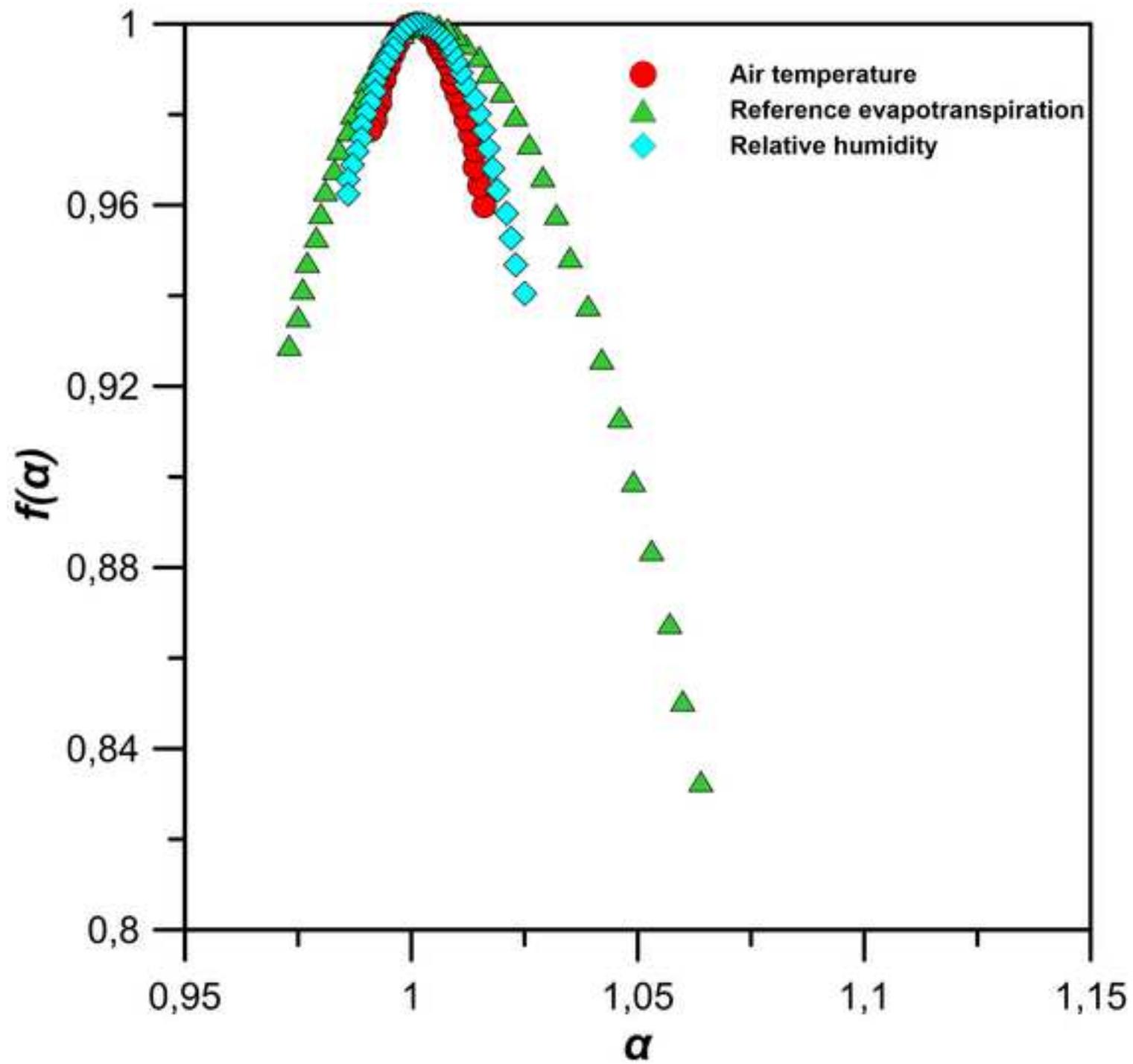

**fig. 6.**

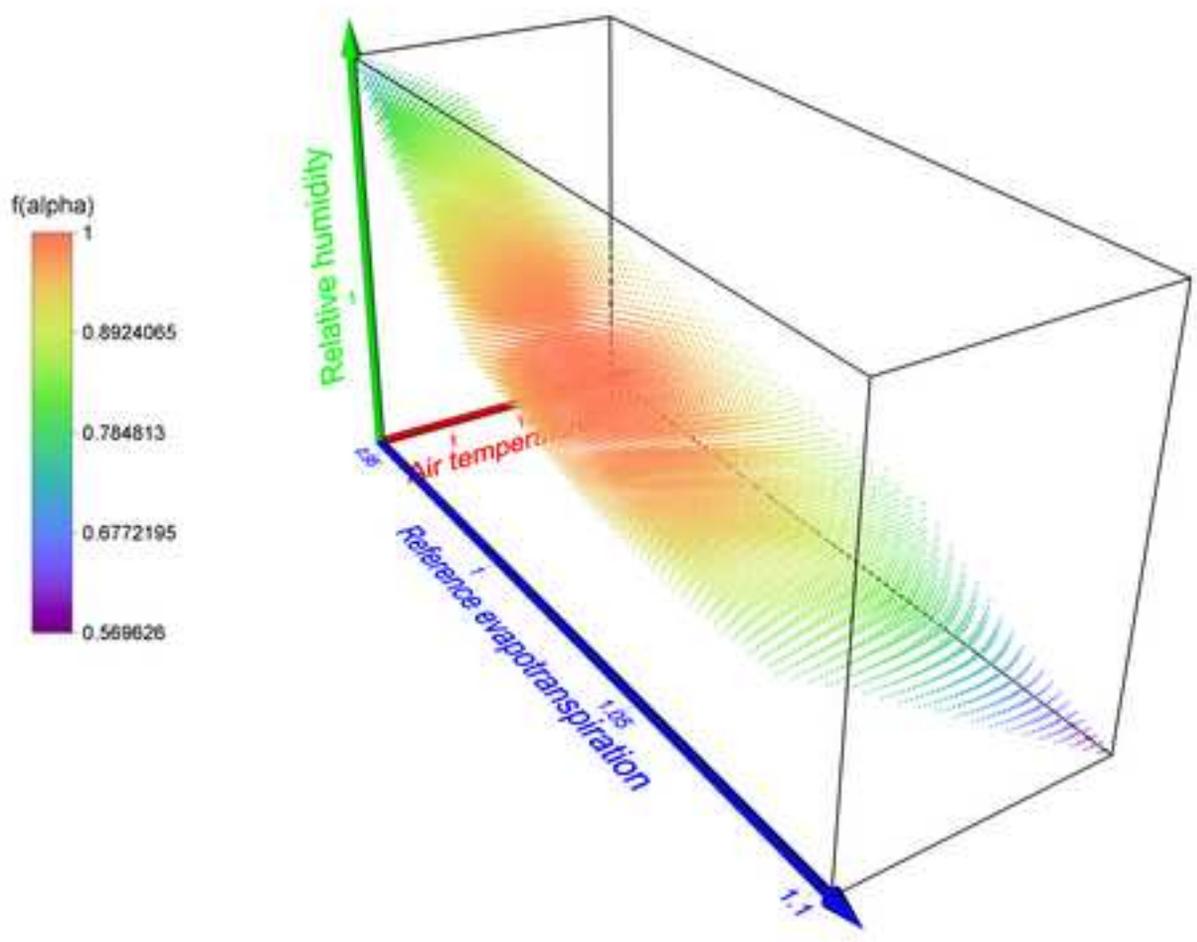

**fig.7.**

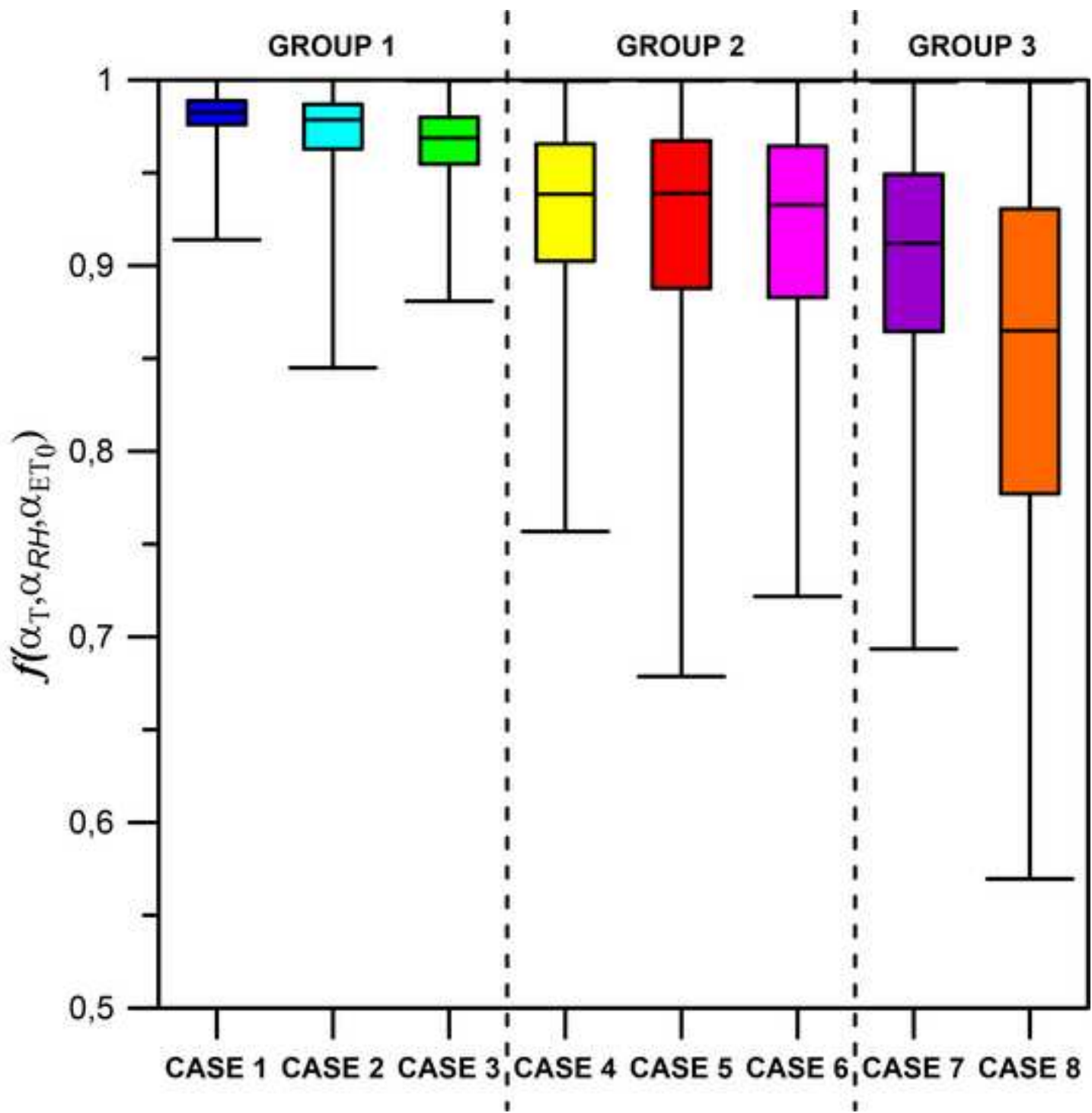

**fig.8.**

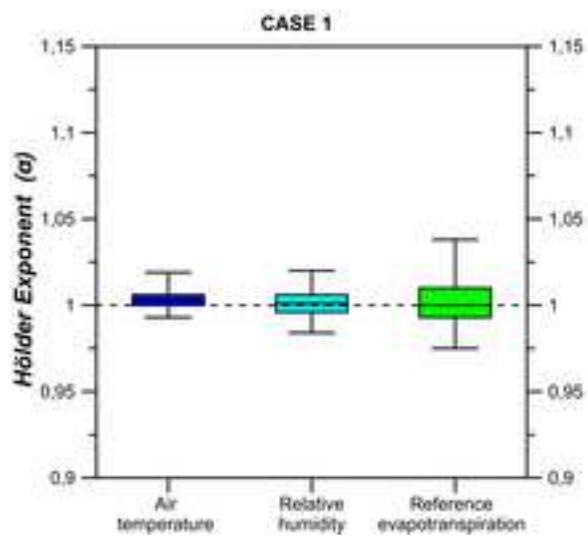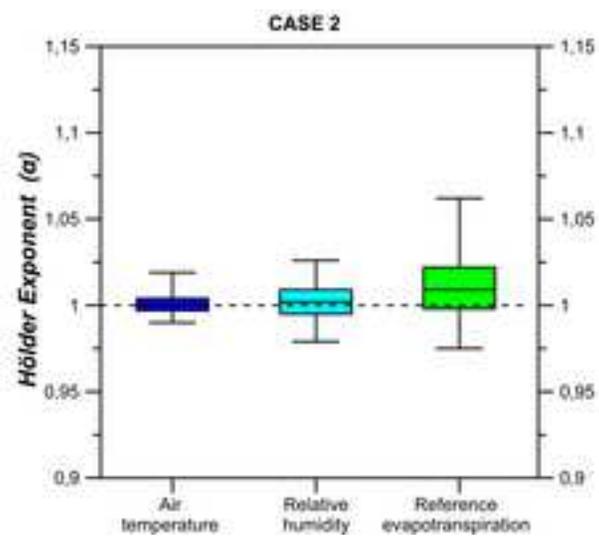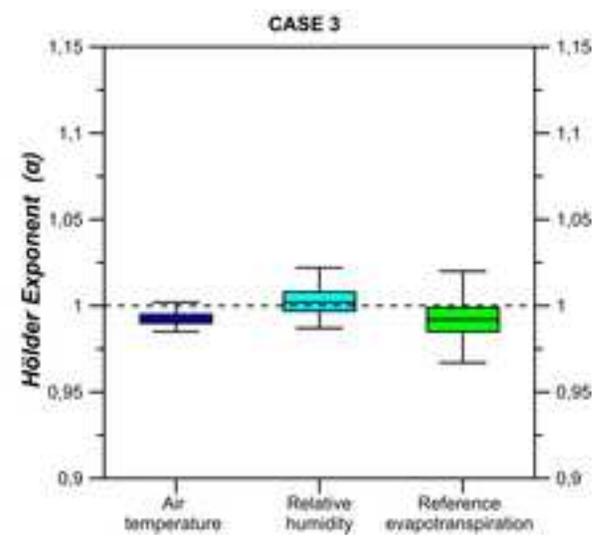

**fig.9.**

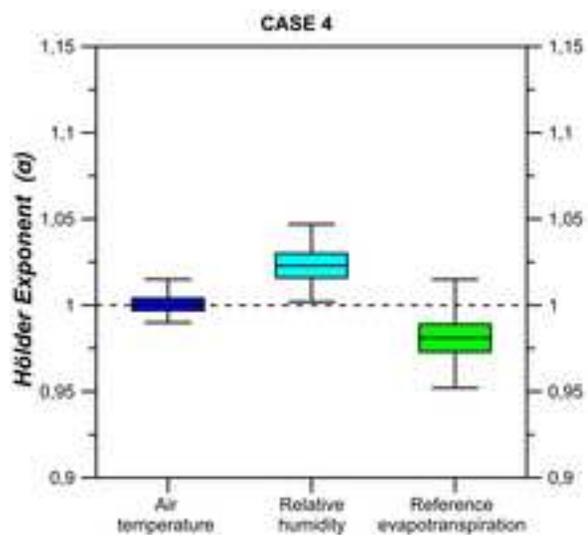 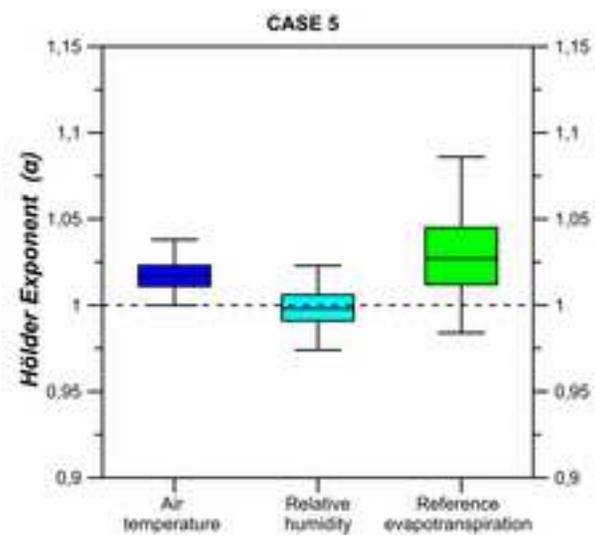 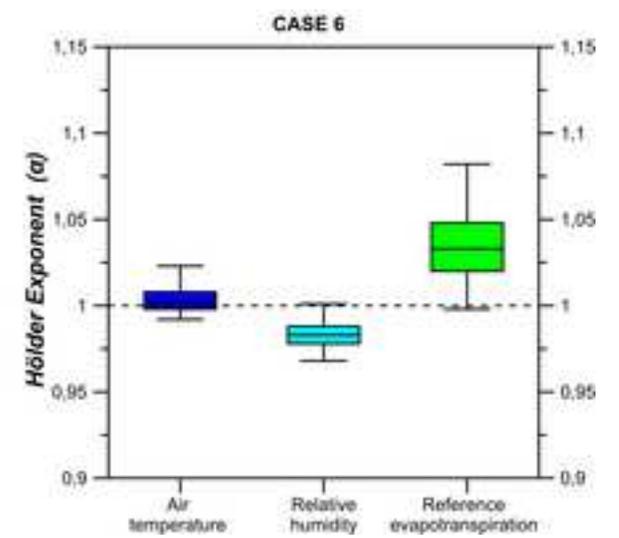

**fig.10.**

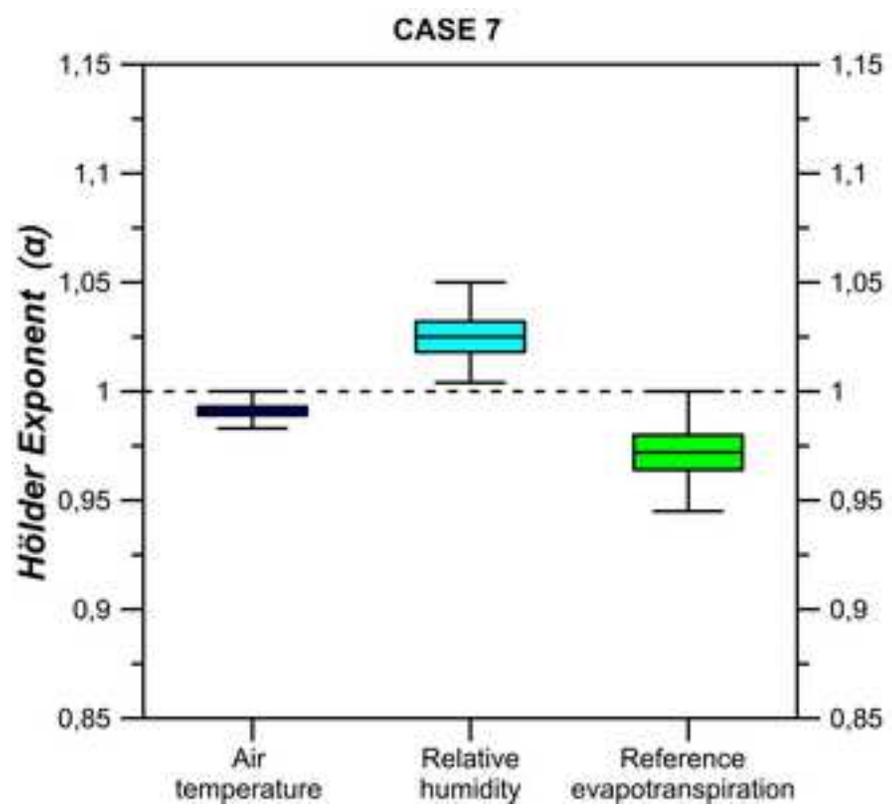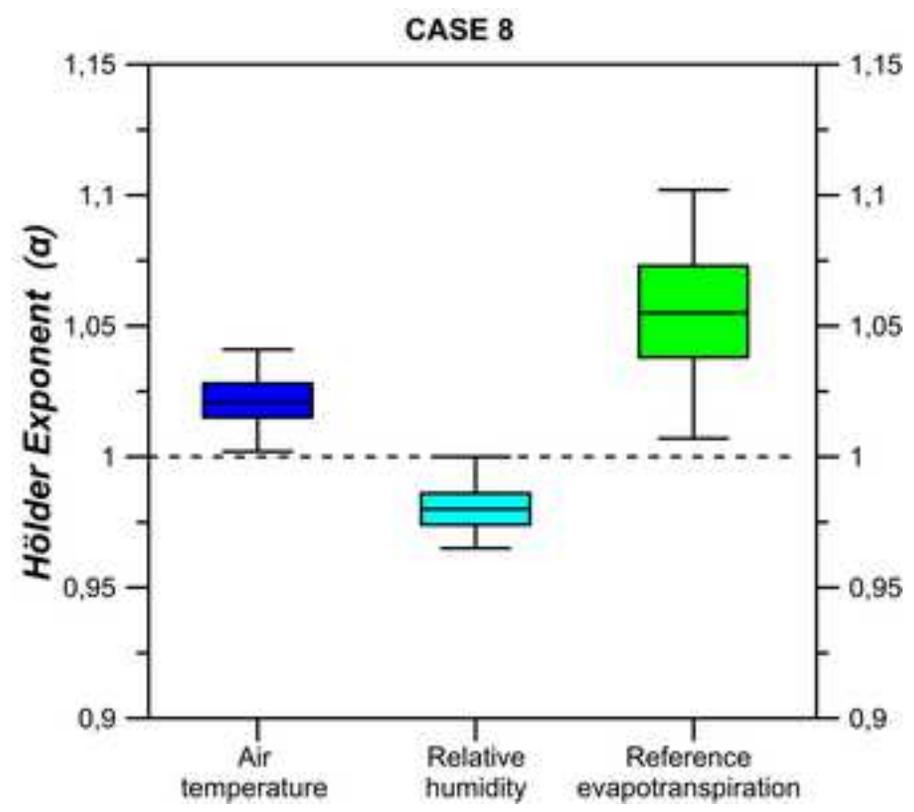